\documentclass[review]{elsarticle}
\usepackage{amsmath}
\usepackage{lineno,hyperref}
\modulolinenumbers[205]
\journal{Journal of \LaTeX\ Templates}
\usepackage{comment}
\usepackage{setspace}
\usepackage{hyperref}
\renewcommand{\baselinestretch}{1.1} 
\usepackage{geometry}
\usepackage{color}
\usepackage[normalem]{ulem}
\newgeometry{vmargin={21mm}, hmargin={20mm,20mm}}

\begin{document}

\begin{frontmatter}
\title{Spatiotemporal instabilities and pattern formation in systems of diffusively coupled Izhikevich neurons}
\tnotetext[]{Corresponding author\\ Email address: arghamondalb1@gmail.com}
\author{ Argha Mondal$^{1, 2, ^*}$, Chittaranjan Hens$^{3}$, Arnab Mondal$^{4}$,  Chris G. Antonopoulos$^2$}
\address{$^{1}$ School of Engineering, Amrita Vishwa Vidyapeetham, Amritapuri, Kollam 690525, India\\
	$^2$ Department of Mathematical Sciences, University of Essex, Wivenhoe Park, UK\\
$^{3}$ Physics and Applied Mathematics Unit, Indian Statistical Institute Kolkata, WB, India\\
$^{4}$ Department of Mathematics and Computing, Indian Institute of Technology (Indian School of Mines), Dhanbad 826004, India 
}

%
%
%
%

\begin{abstract}
Neurons are often connected, spatially and temporally, in phenomenal ways that promote wave propagation. Therefore, it is essential to analyze the emergent spatiotemporal patterns to understand the working mechanism of brain  activity, especially in cortical areas. Here, we present an explicit mathematical analysis, corroborated by numerical results, to identify and investigate the spatiotemporal, non-uniform, patterns that emerge due to instability in an extended homogeneous 2D spatial domain, using the excitable Izhikevich neuron model. We examine diffusive instability and perform bifurcation and fixed-point analyses to characterize the patterns and their stability. Then, we derive analytically the amplitude equations that establish the activities of reaction-diffusion structures. We report on the emergence of diverse spatial structures including hexagonal and mixed-type patterns by providing a systematic mathematical approach, including variations in correlated oscillations, pattern variations and amplitude fluctuations. Our work shows that the emergence of spatiotemporal behavior, commonly found in excitable systems, has the potential to contribute significantly to the study of diffusively-coupled biophysical systems at large.
\end{abstract}


\end{frontmatter}


\section{Introduction}

The spatially structured patterns associated to wave propagation in coupled neural populations can emerge due to high-low amplitude oscillations and variations in the firing activities of single neurons in different network topologies. One of the key challenges is to analyze the network dynamics and characteristics \cite{ermentrout1998neural,izhikevich2007dynamical,keane2015propagating,townsend2018detection,song2018classification}. The emergence of diverse spatial patterns is often found to be associated with self-organization and competition within neurons in a network. This may reflect the dynamics of signal processing and functional connectivity in different brain areas \cite{wu2008propagating,huang2010spiral,meier2015bursting}. Thus, it is important to derive the spatial characteristics of the electrical activity, investigate quantitatively and classify the various patterns observed \cite{schiff2007dynamical,yang2006turing,jiang2015formation,townsend2015emergence}. During wave propagation, single neurons are slightly depolarized, however the probability of firing action potentials or generating high-amplitude oscillations with spikes often increases depending on the synaptic coupling strengths. These membrane currents depolarize the cell which produces a voltage signal, composed of spikes. When a large-size network becomes coherently depolarized during wave propagation, the local transmission between interneurons in the network may be increased. As a result, various spatiotemporal patterns such as spirals, traveling waves and hexagonal, mixed-like patterns can be observed at different levels of activity \cite{yang2006turing,tikhomirova2007nonlinear,chen2006stability,mondal2019diffusion,ma2010transition}. Spatial patterns can also occur in Fitzhugh-Nagumo systems with spatial interactions \cite{ambrosio2012synchronization,ambrosio2019large,carletti2020turing,kuznetsov2017pattern,iqbal2017pattern}. Wave propagation of neural activity is often observed when a population of cortical cells is working during sensory or motor event processing \cite{huang2010spiral}.

Here, we use a systematic, mathematical analysis based on amplitude equations \cite{ipsen2000amplitude,zhao2014turing} to investigate the formation of coherent structures of patterns. In addition, rigorous numerical simulations are performed to identify the diversity in wavy patterns. These patterns emerge due to instabilities in a homogeneous medium (i.e., in a 2D spatial domain) in a reaction-diffusion system of coupled Izhikevich neurons.

In our work, we present a comprehensive analysis to study the onset of pattern formation and the dominant effects. Our work reveals how the analytical method can be used to identify and analyze the spatiotemporal characteristics that can help us in understanding the working mechanism of the collective dynamics. The existence and stability of the spatial patterns are explored analytically with the amplitude equations near the Turing bifurcation points. 
In \cite{mondal2019diffusion}, the authors studied the spatiotemporal dynamics of pattern formation in a coupled 2D Morris-Lecar neural system and investigated the nonlinear responses of an excitable, conductance-based, neuronal cable. They validated their numerical results showing they are in good agreement with their analytical results using amplitude equations and multiple-scale analysis, the same approach we adopt here. Furthermore, they studied the formation of spatiotemporal patterns in non-Turing regimes using the 2D Morris-Lecar  model, whereas here we use diffusively coupled Izhikevich neurons to study the formation of spatiotemporal patterns in Turing regimes, following the same methodology. This method presents a mathematical framework in the extended coupled spatiotemporal systems to establish variety of structurally different patterns with the modulation and stability analysis that are often found in excitable coupled biophysical systems. Thus, our present work and the work in \cite{mondal2019diffusion}, show clearly that the analytical approach used in our study here can be generalized to study analytically the spatiotemporal instabilities and pattern formation in other biophysically excitable models, including in Turing and non-Turing regimes.

Particularly, in our work we explore the fluctuations and local nonlinear excitations of diffusively coupled Izhikevich neurons \cite{izhikevich2003simple,izhikevich2004model,teka2018spiking} given by a system of partial differential equations (PDEs). Performing a detailed bifurcation analysis, we investigate the quiescent and oscillatory regimes and explore the influence of coupling strengths on the system, which undergoes a diffusion-driven instability, known as Turing instability \cite{tang2015bifurcation,tang2020chemotaxis,turing1990chemical,ouyang1991transition,kondo2010reaction}. We study how bursting regimes of a single excitable system can emerge in a diffusively coupled system. The patterns associated with bursting range from regular stable hexagons to distorted hexagons \cite{iqbal2017pattern,sain2000instabilities,yuan2013spatial,zhang2014spatio}. To study the modulation and stability of wave patterns (i.e., hexagons), we consider the phase and amplitude of the oscillatory dynamics. We show that around the Turing bifurcation points of the dynamics, spatiotemporal patterns emerge, that are commonly found in extended biophysical systems \cite{mondal2019diffusion,mhatre2012grid}. Thus, our work has the potential to contribute significantly to the study of diffusively-coupled biophysical systems at large.

\section{The excitable single-neuron Izhikevich model}

This work focuses on the complex dynamics of the Izhikevich model that describes the electrical activity of a single neuron for a range of parameter values. The model exhibits spike generation and a discontinuous resetting process following the spikes. Particularly, its time evolution is described by the set of ordinary differential equations \cite{izhikevich2003simple,izhikevich2004model}
\begin{equation}
\begin{array}{l}
\frac{dv}{dt}=0.04v^2+5v+140-u+I = f_1(v,u),\\
\frac{du}{dt}=a(bv-u) = f_2(v,u),
\end{array}
\label{model}
\end{equation}
where $v$ is the membrane voltage and $u$ the recovery variable that measures the activation of K$^+$ and inactivation of Na$^+$ ionic currents. When the voltage, $v$ reaches its peak value, the following relation is applied after the spike resetting constraint: if $v \ge {v_{peak}} = 30$, then $v \leftarrow c$ and $u \leftarrow u + d$. The resting potential depends on the parameter $b$ that indicates the sensitivity of $u$ to the subthreshold fluctuations of the voltage and, $a$ measures the timescale of the recovery variable, $u$. The parameters $c$ and $d$ control the after-spike reset values of $v$ and $u$, respectively. The function $(0.04v^2 + 5v + 140)$ was derived using the spike initiation dynamics of a cortical cell. The initial conditions are set to $v=-63$ and $u=bv$, respectively \cite{izhikevich2003simple,izhikevich2004model} and the stimulus currents are injected to the neuron with the variable, $I$.

To study the behavior of the Izhikevich model \eqref{model} in the absense of diffusion for different parameter sets and $I$ values, we perturb it around the fixed point $E=(v_0,u_0)$ and compute the corresponding Jacobian matrix, $J$. The equilibrium point $E$ can be deduced by solving the set of equations $f_1(v_0,u_0)=0$ and $f_2(v_0,u_0)=0$ simultaneously for $v_0$ and $u_0$, which gives $v_0=\frac{-(5-b) \pm \sqrt{(5-b)^2-0.16(140+I)}}{0.08}$ and $u_0=b v_0$. The Jacobian of system \eqref{model} computed at $E=(v_0,u_0)$ is given by
$J = \left( {\begin{array}{*{20}{c}}
	{{a_{11}}}&{{a_{12}}}\\
	{{a_{21}}}&{{a_{22}}}
	\end{array}} \right)$,
where $a_{11}=0.08v_0+5$, $a_{12}=-1$, $a_{21}=ab$ and $a_{22}=-a$. The fixed point, $E$ is locally asymptotically stable if, trace$(J)=a_{11}+a_{22}<0$ and $\det(J)=a_{11}a_{22}-a_{12}a_{21}>0$.

\begin{figure}[!ht]
\centering
\includegraphics[width=16cm,height=9.1cm]{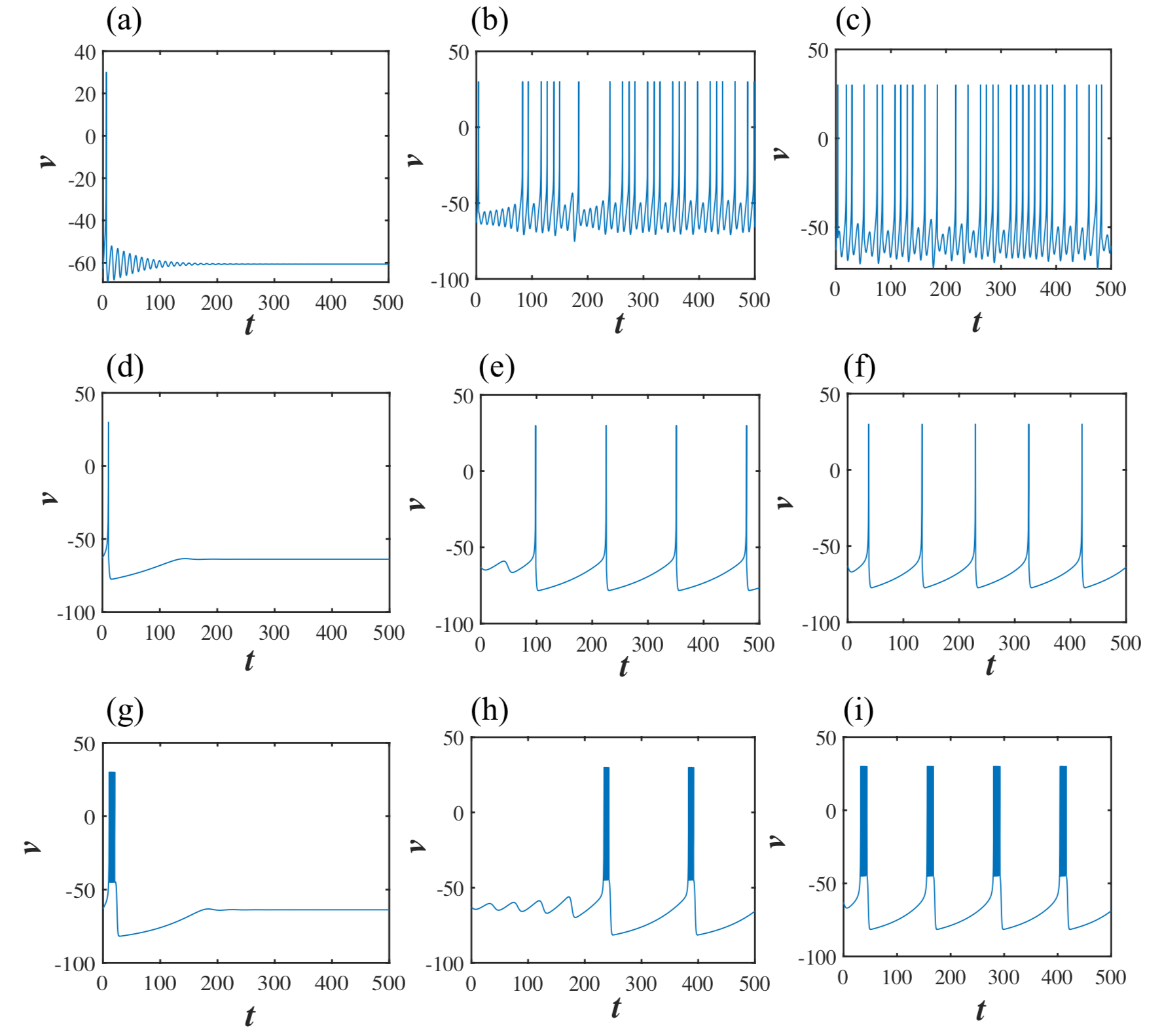}
\caption{Temporal plots of the membrane voltage $v$ in system \eqref{model}, where panels (a)-(c) are for $I=-105.1$, $-103.9, -103$ (set I), (d)-(f) for $I=80$, 78.8, 78 (set II) and (g)-(i) for $I=80$, 78.9, 78 (set III).}\label{time}
\end{figure}

In the following, we consider specific types of biophysically plausible parameter sets that give rise to Hopf bifurcations by varying $I$. This amounts to finding spatial patterns using the method of amplitude equations \cite{ipsen2000amplitude,zhao2014turing} and identifying the onset of Hopf bifurcations. The parameter sets we consider are the following: set I with $a=0.2$, $b= 2$, $c= -56$, $d= -16$, $I\in [-105.1,-103]$, set II with $a=-0.02$, $b= -1$, $c= -60$, $d= 8$, $I\in [78,80]$ and set III with $a=-0.026$, $b= -1$, $c= -45$, $d= 0$, $I\in [78,80]$. System \eqref{model} produces firings for these parameter sets with varying $I$, ranging from quiescent to oscillatory behavior through chaotic spiking (Fig. \ref{time} (b), (c)), phasic to tonic spiking (Fig. \ref{time} (d)-(f)) and phasic bursting to regular bursting shown in Fig. \ref{time} (g)-(i) \cite{izhikevich2003simple,izhikevich2004model}. The numerical simulations were performed using the fourth-order Runge-Kutta method with time step $\delta t=0.01$. We have also checked that the results with a smaller time step do not show any significant differences.

\subsection{Bifurcation analysis}

\begin{figure}[!h]
\centering
\includegraphics[width=16cm,height=5cm]{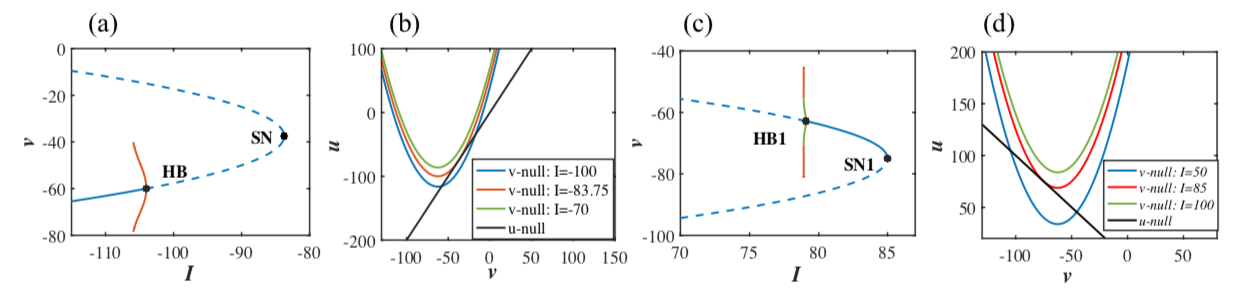}
\caption{Bifurcation diagrams of system \eqref{model}, where $I$ is the bifurcation parameter. Panels (a) and (b) are for parameters in set I and (c), (d) for parameters in set III. Thick and dashed-blue curves represent stable and unstable branches, i.e., stability analysis of the fixed points, green and red curves represent stable and unstable limit cycles. Panels (b), (d) show the nullclines for the emergence of saddle node bifurcations for three different values of $I$. We note that system \eqref{model} undergoes the same type of bifurcation for parameters in set II, and hence a bifurcation plot for this parameter set is not shown. Note also that HB and HB1 stand for Hopf-bifurcation points and SN and SN1 for saddle-node bifurcation points.}\label{bifurcation}
\end{figure}

The bifurcation analysis of system \eqref{model} was performed using MatCont \cite{MatCont_paper} in Matlab, varying the bifurcation parameter $I$. At lower current stimulus $I<-83.75$, the system has two fixed points which collide at $I=-83.75$ (see Fig. \ref{bifurcation}(a), (b), where SN stands for saddle-node bifurcation) and vanish for $I>-83.75$. The system changes its stability and becomes unstable at $I=-104$, where an unstable limit cycle emerges, shown in Fig. \ref{bifurcation}(a), denoted by the point HB for Hopf bifurcation. Particularly, a subcritical Hopf bifurcation (HB) appears at $I=-104$. In Fig. \ref{bifurcation}(a), thick and dashed blue lines represent the quiescent and oscillatory regions, respectively, and the red line, an unstable limit cycle.

In the following, we verify our numerical results performing an analytical study. Particularly, the complex eigenvalues of $J$ computed at $E=(v_0,u_0)$ are given by $\gamma (I),\,\,\overline {\gamma (I)} = \alpha (I) \pm i\beta (I)$. Suppose that, for a certain value $I=I_0$, the following conditions are met: (i) $\alpha(I_0)=0$, (ii) $\beta(I_0)=\beta^*\neq0$ and (iii) ${\left. {\frac{{d\alpha \left( I \right)}}{{dI}}} \right|_{I = I_0}} = \alpha^* \ne 0$. Then, system \eqref{model} undergoes a Hopf bifurcation at $I=I_0$. Conditions (i) and (ii) are known as non-hyperbolicity conditions and condition (iii) as the transversality condition. To verify analytically the Hopf bifurcation varying $I$ for parameters in set I, we obtain $\alpha (I)=0.04(60+v_0)$ and $\beta (I)=0.04\sqrt{-v_0^2-130v_0-3975}$, where $v_0=2.5\big(-15\pm\sqrt{-335-4I}\big)$. To show the existence of a Hopf bifurcation, we find the value of $I=I_0$ at which conditions (i) - (iii) are satisfied. Thus, solving the equation $\alpha(I)=0$ for $I$, we obtain $I=I_0=-104$. Then, from condition (i) we have $\alpha(-104)=0$, from (ii) $\beta \left( {-104} \right) = 0.6 \ne 0$ and from (iii) ${\left. {\frac{{d\alpha \left( I \right)}}{{dI}}} \right|_{I = -104}} = 0.0222 \ne 0$. Thus, we conclude that system \eqref{model} undergoes a Hopf bifurcation at $I=-104$, which is in good agreement with the numerical simulations. The system undergoes a super critical Hopf bifurcation at $I=79.070775$ and a saddle-node bifurcation for parameters in set III (see Fig. \ref{bifurcation}(c), (d)). Interestingly, it undergoes the same type of bifurcation for parameters in set II, and hence a bifurcation plot for parameters in set II is not shown in Fig. \ref{bifurcation}.

\section{Diffusively coupled Izhikevich models on a 2D spatial domain}

In the following, we consider the system of diffusively coupled Izhikevich neurons 
\begin{equation}
\begin{array}{l}
\frac{{\partial v}}{{\partial t}} = f_1(v,u)+D_{11}\Big(\frac{\partial ^2v}{\partial x^2}+\frac{\partial ^2v}{\partial y^2}\Big),\\
\frac{{\partial u}}{{\partial t}}=f_2(v,u)+D_{22}\Big(\frac{\partial ^2u}{\partial x^2}+\frac{\partial ^2u}{\partial y^2}\Big),
\end{array}
\label{2Dmodel}
\end{equation}
on a 2D spatial domain $\Omega$, extending the single deterministic Izhikevich model \eqref{model}. In this framework, $v\equiv v(x,y,t)$, $u\equiv u(x,y,t)$ and $t$, $x$ and $y$ represent the time, $x$ and $y$ spatial coordinates, respectively. Therefore, the solution $(v,u)$ represents the activity of the Izhikevich neurons in $x$ and $y$ at time $t$. Parameters $D_{11}$ and $D_{22}>0$ quantify the diffusion coupling strengths. The initial conditions are given by the known functions $v(x,y,t = 0) $ and $u(x,y,t = 0) $ for $x,y\in \Omega$, where $\Omega$ is the bounded-square or spatial domain. Moreover, we consider zero-flux boundary conditions $\frac{{\partial v}}{{\partial n}}=\frac{{\partial u}}{{\partial n}}=0$, $x,y \in \partial \Omega $ and $t > 0$, where $n$ is the outward normal to the boundary of $\Omega$, denoted by $\partial \Omega $. The zero-flux boundary conditions imply that the cell membranes are impermeable to ionic movements at the boundaries \cite{meier2015bursting}.

For diffusive systems such as system \eqref{2Dmodel}, the Turing condition \cite{tang2015bifurcation,turing1990chemical} implies that its steady-state solution is unstable around a fixed point and stable for the system with no diffusion, such as system \eqref{model}. Next, we perturb system \eqref{2Dmodel} around the uniform steady-state condition and linearize it around the nontrivial fixed point, obtaining the characteristic equation, assuming the particular solution $\left( {\begin{array}{*{20}{c}}
	v\\
	u
	\end{array}} \right) = \left( {\begin{array}{*{20}{c}}
	{{v_0}}\\
	{{u_0}}
	\end{array}} \right) + \varepsilon \left( {\begin{array}{*{20}{c}}
	{{v_k}}\\
	{{u_k}}
	\end{array}} \right){e^{\lambda _k t + ikr}} + c.c. + o\left( {{\varepsilon ^2}} \right)$, where $c.c$. stands for complex conjugate, $\lambda _k$ is the wave length and $k$ the wave number in the $r$ direction, where $r=(x,y)$ is the directional vector in two dimensions. The Jacobian matrix, $J_D$, of Eq. \eqref{2Dmodel} computed at the fixed point $E=(v_0,u_0)$ is then given by
$J_D = J-k^2D=\left( {\begin{array}{*{20}{c}}
	{{a_{11}-D_{11}k^2}}&{{a_{12}}}\\
	{{a_{21}}}&{{a_{22}-D_{22}k^2}}
	\end{array}} \right),$
where 
$D = \left( {\begin{array}{*{20}{c}}
	{{D_{11}}}&{{0}}\\
	{{0}}&{{D_{22}}}
	\end{array}} \right)$.
The stability conditions of system \eqref{2Dmodel} at $E=(v_0,u_0)$ are given by trace$(J_D)=a_{11}+a_{22}-D_{11}k^2-D_{22}k^2<0$ and $\det(J_D)=(a_{11}-D_{11}k^2)(a_{22}-D_{22}k^2)-a_{12}a_{21}>0$. The eigenvalues of $J_D$ are given by $\lambda = \frac{1}{2}\Big(\mbox{trace}(J_D) \pm \sqrt{\mbox{trace}^2(J_D)-4\det(J_D)}\Big)$, which is also known as the dispersion relation. If we consider the stable fixed point of system \eqref{model}, then trace$(J_D)<0$ as $a_{11}+a_{22}<0$, $D>0$ and $k^2>0$. Thus, the conditions for the Turing instability become $\mbox{trace}(J)=a_{11}+a_{22}<0$, $\det(J)=a_{11}a_{22}-a_{12}a_{21}>0$ and $\det(J_D)=\big(a_{11}-D_{11}k^2\big)\big(a_{22}-D_{22}k^2\big)-a_{12}a_{21}<0$.

\begin{figure}[!ht]
\centering
\includegraphics[width=18cm,height=11cm]{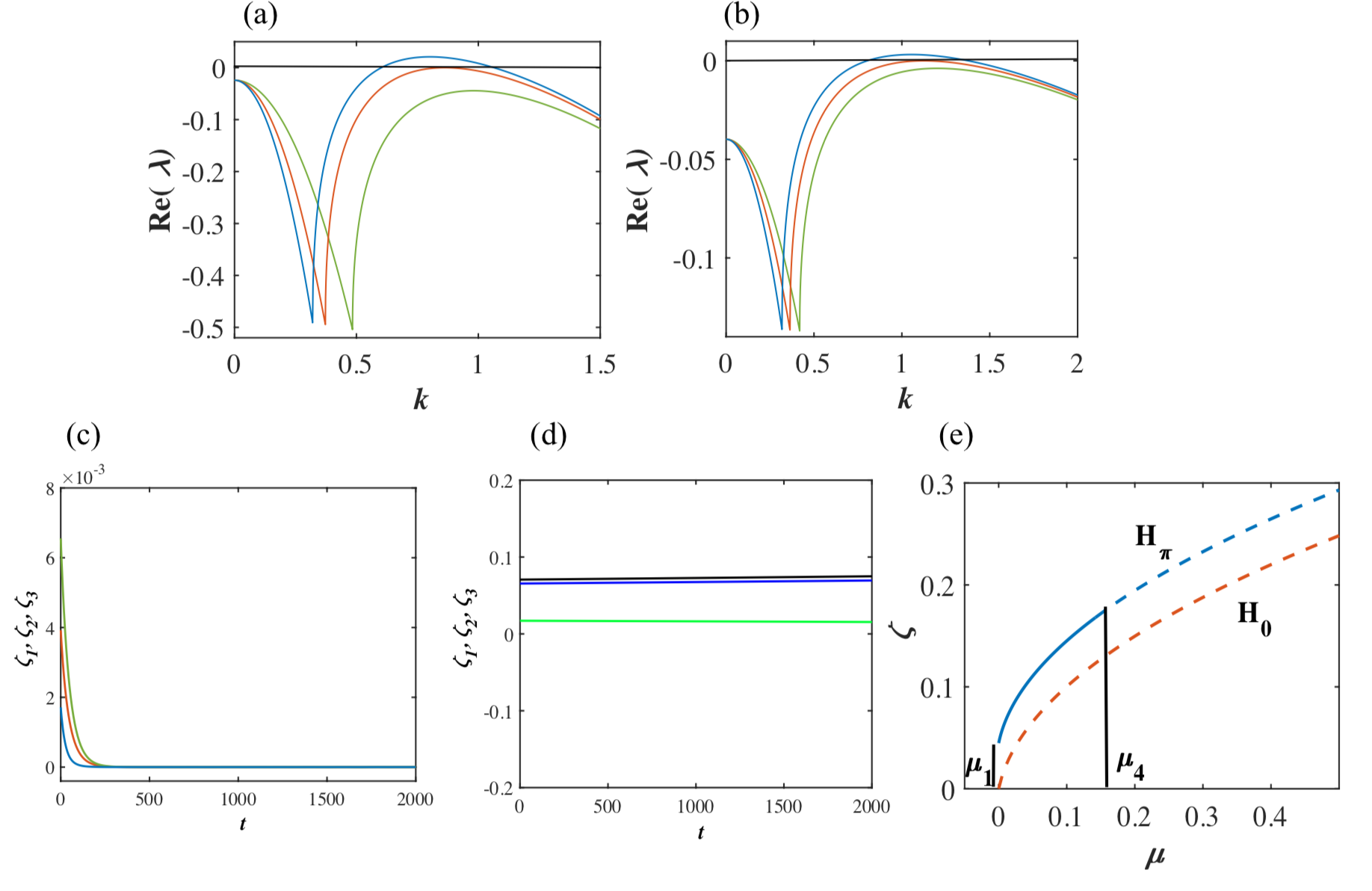}
\caption{Plots of the dispersion relation for system \eqref{2Dmodel} for various couplings: (a)  $I=-105.1$, $D_{11}=0.1$, where the green, red and blue curves correspond to $D_{22}=4$, 6.68117 and 9 (set I), (b)  $I=80$, $D_{22}=0.01$, where the green, red and blue lines correspond to $D_{11}=1.1$, 1.45612 and 1.9 (set III), respectively. Panels (c) and (d) show plots of Eq. \eqref{finalampli} for (c) $I=-105.1$, $D_{11}=0.1$, $D_{22}=9$ (set I) and (d) $I=80$, $D_{11}=1.46$, $D_{22}=0.01$ (set III). Note that $\lambda$ are the eigenvalues of the Jacobian matrix $J_D$, thus Re$(\lambda)$ are their real parts and that $\zeta_1$, $\zeta_2$, $\zeta_3$ are the modes discussed in Sec. \ref{subesec_as}. (e) Bifurcation diagram of model \eqref{2Dmodel} for parameter set I with $I=-105.1$, $D_{11}=0.1$ and $D_{22}=9$. $H_0$: hexagonal patterns with $\varphi=0$, $H_\pi$: hexagonal patterns with $\varphi=\pi$. Solid and dashed lines indicate stable and unstable states.}
\label{dispersion}
\end{figure}

To appreciate the Turing instability, we have computed numerically and present in Fig. \ref{dispersion}(a), (b) the dispersion relation for various couplings. The function $\det(J_D)$ is quadratic in $k^2$, say $H(k^2)$, which can be written as $ H(k^2)=D_{11}D_{22}k^4 +(-a_{22}D_{11}-a_{11}D_{22})k^2 + \det(J)$. The minimum value of $H(k^2)$ is attained at the critical wave number $k=k_T$, where
\begin{eqnarray}
k_T^2=\frac{a_{22}D_{11}+a_{11}D_{22}}{2D_{11}D_{22}}.
\label{kT}
\end{eqnarray}
Note that, setting $H(k_T^2)=0$ and using Eq. \eqref{kT}, we obtain the critical wave number $k_T=\left( \frac{\det(J)}{\det(D)}\right)^{\frac{1}{4}}$. In the next section, we derive the amplitude equations of system \eqref{2Dmodel} in the vicinity of the Turing bifurcation point, where the system exhibits critical slowing down. Here, we only consider constant diagonal diffusion couplings as this leads to interesting phenomena in excitable spatial systems. In the neighborhood of a Turing bifurcation point, the spatial symmetry of the system is destroyed and a stationary form arises in both time and oscillatory dynamics that can be described by the amplitude equations, discussed in the next section.

\subsection{Amplitude Equations}

The characteristics of different pattern formations around the bifurcation point can be analyzed further by studying the amplitude equations of system \eqref{2Dmodel}. Particularly, these equations can be derived by using a multiple-scale analysis \cite{ipsen2000amplitude,zhao2014turing,yan2020pattern}. Perturbing the system around the fixed point $E=\left( {{v_0},{u_0}} \right)$ by considering $v=\tilde v+{v_0}$ and $u=\tilde u+{u_0}$, we obtain the set of equations
\begin{equation}
\begin{array}{l}
\frac{{\partial \tilde v}}{{\partial t}} = {a_{11}}\tilde v + {a_{12}}\tilde u + 0.04{{\tilde v}^2} + D_{11}{\nabla ^2}\tilde v,\\
\frac{{\partial \tilde u}}{{\partial t}} = {a_{21}}\tilde v + {a_{22}}\tilde u + D_{22}{\nabla ^2}\tilde u,
\end{array}
\label{4.1}
\end{equation}
where $\tilde v$ and $\tilde u$ are small perturbations to $v_0$ and $u_0$, respectively. Equation \eqref{4.1} can be written in the compact form
\begin{equation}
\frac{{\partial P}}{{\partial t}} = LP + N,
\label{4.2}
\end{equation}
where $L$ and $N$ are the linear operator and nonlinear term, respectively, that we discuss below. We Taylor-expand the bifurcation parameter $I$, which is close to the bifurcation threshold, $I_T$, as $I - I_T = \varepsilon {I_1} + {\varepsilon^2}{I_2} + {\varepsilon^3}{I_3} + o\left( {{\varepsilon^4}} \right)$, where $\left| \varepsilon \right| < 1$. We can write $P$ and $N$ as
\begin{align}\label{4.3}
P &= \left( {\begin{array}{*{20}{c}}
	{{\tilde{v}}}\\
	{{\tilde{u}}}
	\end{array}} \right) = \varepsilon \left( {\begin{array}{*{20}{c}}
	{{\tilde{v}_1}}\\
	{{\tilde{u}_1}}
	\end{array}} \right) + {\varepsilon ^2}\left( {\begin{array}{*{20}{c}}
	{{\tilde{v}_2}}\\
	{{\tilde{u}_2}}
	\end{array}} \right) + {\varepsilon ^3}\left( {\begin{array}{*{20}{c}}
	{{\tilde{v}_3}}\\
	{{\tilde{u}_3}}
	\end{array}} \right) + o\left( {{\varepsilon ^3}} \right),\\
N &= {\varepsilon ^2}{n_2} + {\varepsilon ^3}{n_3} + o\left( {{\varepsilon ^3}} \right).
\end{align}
In the last equation, $n_2$ and $n_3$ represent the second and third order of $\varepsilon$ and, are given by ${n_2} = \left( {\begin{array}{*{20}{c}}
	{ 0.04{\tilde{v}_1}^2}\\
	0
	\end{array}} \right)$ and 
${n_3} = \left( {\begin{array}{*{20}{c}}
	{ 0.08\tilde{v}_1\tilde{v}_2}\\
	0
	\end{array}} \right)$, respectively.
The linear operator $L$ can be written in the form 
\begin{equation}
\begin{array}{l}
L = {L_T} + \left( {I - I_T} \right){L_1} + {\left( {I - I_T} \right)^2}{L_2} + o\left( {{{\left( {I - I_T} \right)}^3}} \right) = {L_T} + \left( {I - I_T} \right)M + o\left( {{{\left( {I - I_T} \right)}^2}} \right),
\end{array}
\label{4.5}
\end{equation}
where ${L_i} = \frac{1}{{i!}}\frac{{{\partial ^i}L}}{{\partial {I^i}}},\,i=1,2,\;{L_T} = \left( {\begin{array}{*{20}{c}}
	{{a_{11}^T} + D_{11}{\nabla ^2}}&{{a_{12}^T}}\\
	{{a_{21}^T}}&{{a_{22}^T}+ D_{22}{\nabla ^2}}
	\end{array}} \right)$ and $M = \left( {\begin{array}{*{20}{c}}
	{{m_{11}}}&0\\
	0&0
	\end{array}} \right)$ with $m_{11}=\frac{-0.08}{0.08v_0+5-b}$.
	
Next, we introduce the multiple-scale \cite{zhao2014turing,yan2020pattern} $\frac{\partial }{{\partial t}} = \varepsilon \frac{\partial }{{\partial {T_1}}} + {\varepsilon ^2}\frac{\partial }{{\partial {T_2}}} + o\left( {{\varepsilon ^2}} \right)$, where ${T_1} = \varepsilon t$ and ${T_2} = {\varepsilon ^2}t$. Substituting Eqs. \eqref{4.3} - \eqref{4.5} into Eq. \eqref{4.2}, we obtain 
\begin{equation}
\begin{array}{l}
\left( {\varepsilon \frac{\partial }{{\partial {T_1}}} + {\varepsilon ^2}\frac{\partial }{{\partial {T_2}}}} \right)\left[{\varepsilon \left( {\begin{array}{*{20}{c}}
		{{\tilde{v}_1}}\\
		{{\tilde{u}_1}}
		\end{array}} \right) + {\varepsilon ^2}\left( {\begin{array}{*{20}{c}}
		{{\tilde{v}_2}}\\
		{{\tilde{u}_2}}
		\end{array}} \right) + {\varepsilon ^3}\left( {\begin{array}{*{20}{c}}
		{{\tilde{v}_3}}\\
		{{\tilde{u}_3}}
		\end{array}} \right)} \right] = \\
\left( {{L_T} + \left( {I - I_T} \right)M} \right)\left[{\varepsilon \left( {\begin{array}{*{20}{c}}
		{{\tilde{v}_1}}\\
		{{\tilde{u}_1}}
		\end{array}} \right) + {\varepsilon ^2}\left( {\begin{array}{*{20}{c}}
		{{\tilde{v}_2}}\\
		{{\tilde{u}_2}}
		\end{array}} \right) + {\varepsilon ^3}\left( {\begin{array}{*{20}{c}}
		{{\tilde{v}_3}}\\
		{{\tilde{u}_3}}
		\end{array}} \right)} \right] + {\varepsilon ^2}{n_2} + {\varepsilon ^3}{n_{3.}}
\end{array}
\label{4.6}
\end{equation}
Equating the coefficients of $\varepsilon$, $\varepsilon^2$ and $\varepsilon^3$ in both sides of Eq. \eqref{4.6}, we obtain the equations
\begin{equation}
{L_T}\left( {\begin{array}{*{20}{c}}
	{{\tilde{v}_1}}\\
	{{\tilde{u}_1}}
	\end{array}} \right) = 0,
\label{4.8}
\end{equation}
\begin{equation}
{L_T}\left( {\begin{array}{*{20}{c}}
	{{\tilde{v}_2}}\\
	{{\tilde{u}_2}}
	\end{array}} \right) = \frac{\partial }{{\partial {T_1}}}\left( {\begin{array}{*{20}{c}}
	{{\tilde{v}_1}}\\
	{{\tilde{u}_1}}
	\end{array}} \right) - {I_1}M\left( {\begin{array}{*{20}{c}}
	{{\tilde{v}_1}}\\
	{{\tilde{u}_1}}
	\end{array}} \right) - {n_2} \buildrel \Delta \over = \left( {\begin{array}{*{20}{c}}
	{{F_{\tilde{v}}}}\\
	{{F_{\tilde{u}}}}
	\end{array}} \right),
\label{4.9}
\end{equation}
and
\begin{equation}
{L_T}\left( {\begin{array}{*{20}{c}}
	{{\tilde{v}_3}}\\
	{{\tilde{u}_3}}
	\end{array}} \right) = \frac{\partial }{{\partial {T_1}}}\left( {\begin{array}{*{20}{c}}
	{{\tilde{v}_2}}\\
	{{\tilde{u}_2}}
	\end{array}} \right) + \frac{\partial }{{\partial {T_2}}}\left( {\begin{array}{*{20}{c}}
	{{\tilde{v}_1}}\\
	{{\tilde{u}_1}}
	\end{array}} \right) - {I_1}M\left( {\begin{array}{*{20}{c}}
	{{\tilde{v}_2}}\\
	{{\tilde{u}_2}}
	\end{array}} \right) - {I_2}M\left( {\begin{array}{*{20}{c}}
	{{\tilde{v}_1}}\\
	{{\tilde{u}_1}}
	\end{array}} \right) - {n_3}.
\end{equation}
Solving Eq. \eqref{4.8}, we obtain ${\tilde{v}_1} = \frac{{a_{11}^T}D_{22}-{a_{22}^T}D_{11}}{2D_{11}{a_{21}^T}} = {f}$ and ${\tilde{u}_1} = 1$. We can write 
\begin{equation*}
\left( {\begin{array}{*{20}{c}}
	{{\tilde{v}_1}}\\
	{{\tilde{u}_1}}
	\end{array}} \right) = \left( {\begin{array}{*{20}{c}}
	{{f}}\\
	1
	\end{array}} \right) {\sum\limits_{j = 1}^3 {{\Theta_j}} e^{i{k_j}.r} + c.c},
\end{equation*}
where $(\tilde{v}_1,\tilde{u}_1)$ is the linear combination of the eigenvectors that corresponds to the zero eigenvalue of the linear operator $L_T$. In this context, $\Theta_j$ denotes the amplitude of the mode $e^{i{k_j}.r}$. We use the Fredholm solvability condition \cite{yuan2013spatial,yan2020pattern} to find the nontrivial solution of Eq. \eqref{4.9}. The right hand-side of Eq. \eqref{4.9} and the eigenvectors corresponding to the zero eigenvalue of $L_{T}^{+}$ must be orthogonal. $L_{T}^{+}$ indicates the adjoint operator of $L_T$ and the eigenvectors of $L_{T}^{+}$ are given by
$\left( {\begin{array}{*{20}{c}}
	{{1}}\\
	g
	\end{array}} \right) e^{ik_{j}r}$, where $j=1,2,3$ and $g=\frac{{a_{22}^T}D_{11}-{a_{11}^T}D_{22}}{2D_{22}{a_{21}^T}}$. Applying the orthogonality condition, $\left( {\begin{array}{*{20}{c}}
	{{1}}\,\,\,\,
	{{g}}
	\end{array}} \right)\left( {\begin{array}{*{20}{c}}
	{{F_{\tilde{p}}^{i}}}\\
	F_{\tilde{q}}^{i}
	\end{array}} \right)=0$, we obtain the equations
\begin{equation*}
\begin{array}{l}
\left( {f+g} \right)\frac{{\partial {\Theta_1}}}{{\partial {T_1}}} = {I_1}{{m_{11}}{f}} {\Theta_1} - 2{h_1}{{\bar \Theta}_2}{{\bar \Theta}_3},\\
\left( {f+g} \right)\frac{{\partial {\Theta_2}}}{{\partial {T_1}}} = {I_1} {{m_{11}}{f}} {\Theta_2} - 2{h_1}{{\bar \Theta}_1}{{\bar \Theta}_3},\\
\left( {f+g} \right)\frac{{\partial {\Theta_3}}}{{\partial {T_1}}} = {I_1} {{m_{11}}{f}} {\Theta_3} - 2{h_1}{{\bar \Theta}_1}{{\bar \Theta}_2},
\end{array}
\end{equation*}
where $F_{\tilde{v}}^{i}$, $F_{\tilde{u}}^{i}$ are the coefficients of $e^{ik_i.r}$ in $F_{\tilde{v}}$, $F_{\tilde{u}}$ and ${h_1} = -0.04{f_1}^2$.
Solving Eq. \eqref{4.9}, we obtain
\begin{equation}
\begin{array}{l}
\left( {\begin{array}{*{20}{c}}
	{{\tilde{v}_2}}\\
	{{\tilde{u}_2}}
	\end{array}} \right) = \left( {\begin{array}{*{20}{c}}
	{{V_0}}\\
	{{U_0}}
	\end{array}} \right) + \sum\limits_{j = 1}^3 {\left( {\begin{array}{*{20}{c}}
		{{V_j}}\\
		{{U_j}}
		\end{array}} \right)e^{i{k_j}.r}} + \sum\limits_{j = 1}^3 {\left( {\begin{array}{*{20}{c}}
		{{V_{jj}}}\\
		{{U_{jj}}}
		\end{array}} \right)e^{i2{k_j}.r}} + \left( {\begin{array}{*{20}{c}}
	{{V_{12}}}\\
	{{U_{12}}}
	\end{array}} \right)e^{i({k_1} - {k_2}).r} + \\
\left( {\begin{array}{*{20}{c}}
	{{V_{23}}}\\
	{{U_{23}}}
	\end{array}} \right)e^{i({k_2} - {k_3}).r} + \left( {\begin{array}{*{20}{c}}
	{{V_{31}}}\\
	{{U_{31}}}
	\end{array}} \right)e^{i({k_3} - {k_1}).r} + c.c,
\end{array}
\label{4.12}
\end{equation}
where, the coefficients are given in the Appendix. Again, using the orthogonality condition, we get
\begin{equation}
\begin{array}{l}
\vspace{0.2cm}
\left( {f+g} \right)\left( {\frac{{\partial {U_1}}}{{\partial {T_1}}} + \frac{{\partial {\Theta_1}}}{{\partial {T_2}}}} \right) = {m_{11}}f\left( {{I_1}{U_1} + {I_2}{\Theta_1}} \right) + H\left( {{{\bar Q}_2}{{\bar \Theta}_3} + {{\bar Q}_3}{{\bar \Theta}_2}} \right) - \,\left[ {{\sigma _1}{{\left| {{\Theta_1}} \right|}^2} + {\sigma _2}\left( {{{\left| {{\Theta_2}} \right|}^2} + {{\left| {{\Theta_3}} \right|}^2}} \right)} \right]{\Theta_1},\\ \vspace{0.2cm}
\left( {f+g} \right)\left( {\frac{{\partial {U_2}}}{{\partial {T_1}}} + \frac{{\partial {\Theta_2}}}{{\partial {T_2}}}} \right) = {m_{11}}f\left( {{I_1}{U_2} + {I_2}{\Theta_2}} \right) + H\left( {{{\bar Q}_1}{{\bar \Theta}_3} + {{\bar Q}_3}{{\bar \Theta}_1}} \right) - \,\left[ {{\sigma _1}{{\left| {{\Theta_1}} \right|}^2} + {\sigma _2}\left( {{{\left| {{\Theta_2}} \right|}^2} + {{\left| {{\Theta_3}} \right|}^2}} \right)} \right]{\Theta_2},\\
\left( {f+g} \right)\left( {\frac{{\partial {U_3}}}{{\partial {T_1}}} + \frac{{\partial {\Theta_3}}}{{\partial {T_2}}}} \right) = {m_{11}}f\left( {{I_1}{U_3} + {I_2}{\Theta_3}} \right) + H\left( {{{\bar Q}_1}{{\bar \Theta}_2} + {{\bar Q}_2}{{\bar \Theta}_1}} \right) - \,\left[ {{\sigma _1}{{\left| {{\Theta_1}} \right|}^2} + {\sigma _2}\left( {{{\left| {{\Theta_2}} \right|}^2} + {{\left| {{\Theta_3}} \right|}^2}} \right)} \right]{\Theta_3}.
\end{array}
\label{4.13}
\end{equation}
Furthermore, the amplitude ${\Lambda_j}$, $j = 1,2,3$ can be expanded as ${\Lambda_j} = \varepsilon {\Theta_j} + {\varepsilon ^2}{U_j} + o({\varepsilon ^3})$. Using the expression for $\Lambda _j$ and Eq. \eqref{4.13}, we obtain the amplitude equations that correspond to $\Lambda_j$,
\begin{equation}
\begin{array}{l}
{\tau _0}\frac{{\partial {\Lambda_1}}}{{\partial t}} = \mu {\Lambda_1} + h{{\bar \Lambda}_2}{{\bar \Lambda}_3} - \left[ {\sigma_1^\prime {{\left| {{\Lambda_1}} \right|}^2} + \sigma_2^\prime \left( {{{\left| {{\Lambda_2}} \right|}^2} + {{\left| {{\Lambda_3}} \right|}^2}} \right)} \right]{\Lambda_1},\\
{\tau _0}\frac{{\partial {\Lambda_2}}}{{\partial t}} = \mu {\Lambda_2} + h{{\bar \Lambda}_1}{{\bar \Lambda}_3} - \left[ {\sigma_1^\prime {{\left| {{\Lambda_2}} \right|}^2} + \sigma_2^\prime \left( {{{\left| {{\Lambda_1}} \right|}^2} + {{\left| {{\Lambda_3}} \right|}^2}} \right)} \right]{\Lambda_2},\\
{\tau _0}\frac{{\partial {\Lambda_3}}}{{\partial t}} = \mu {\Lambda_3} + h{{\bar \Lambda}_1}{{\bar \Lambda}_2} - \left[ {\sigma_1^\prime {{\left| {{\Lambda_3}} \right|}^2} + \sigma_2^\prime \left( {{{\left| {{\Lambda_1}} \right|}^2} + {{\left| {{\Lambda_2}} \right|}^2}} \right)} \right]{\Lambda_3}.
\end{array}
\label{ampli}
\end{equation}
All the parameters involved in Eqs. \eqref{4.13} and \eqref{ampli} are discussed in the Appendix.

\subsection{Amplitude Stability}\label{subesec_as}

Next, we study the stability of the patterns obtained near the bifurcation point. The amplitude $\Lambda_j$ can be decomposed into the mode $\zeta _j = |\Lambda_j|$ and a corresponding phase-angle $\varphi _j$. Then, substituting ${\Lambda_j} = {\zeta _j}e^{i{\varphi _j}}$ into Eq. \eqref{ampli}, and separating the real and imaginary parts, we obtain the following system of ordinary differential equations for the real variables 
\begin{equation}
\begin{array}{l}
{\tau _0}\frac{{\partial \varphi }}{{\partial t}} = - h\frac{{\zeta _1^2\zeta _2^2 + \zeta _1^2\zeta _3^2 + \zeta _2^2\zeta _3^2}}{{{\zeta _1}{\zeta _2}{\zeta _3}}}\sin \varphi, \\ 
{\tau _0}\frac{{\partial {\zeta _1}}}{{\partial t}} = \mu {\zeta _1} + h{\zeta _2}{\zeta _3}\cos \varphi - \sigma_1^\prime\zeta _1^3 - \sigma_2^\prime\big(\zeta _2^2 + \zeta _3^2\big){\zeta _1}, \\ 
{\tau _0}\frac{{\partial {\zeta _2}}}{{\partial t}} = \mu {\zeta _2} + h{\zeta _1}{\zeta _3}\cos \varphi - \sigma_1^\prime\zeta _2^3 - \sigma_2^\prime\big(\zeta _1^2 + \zeta _3^2\big){\zeta _2}, \\ 
{\tau _0}\frac{{\partial {\zeta _3}}}{{\partial t}} = \mu {\zeta _3} + h{\zeta _1}{\zeta _2}\cos \varphi - \sigma_1^\prime\zeta _3^3 - \sigma_2^\prime\big(\zeta _1^2 + \zeta _2^2\big){\zeta _3},
\end{array}
\label{mixed}
\end{equation}
where $\varphi = {\varphi _1} + {\varphi _2} + {\varphi _3}$. Here, we are interested in the stable steady-state. From the first equation in Eq. \eqref{mixed}, we have that $\varphi =0$ or $\pi$, as $h\zeta _i \neq 0$. The state corresponding to $\varphi =0$ and $\varphi = \pi$ is stable when $h>0$ and $h<0$, respectively. Then, the amplitudes of Eq. \eqref{mixed} are given by
\begin{equation}
\begin{array}{l}
{\tau _0}\frac{{\partial {\zeta _1}}}{{\partial t}} = \mu {\zeta _1} + h{\zeta _2}{\zeta _3} - \sigma_1^\prime\zeta _1^3 - \sigma_2^\prime\big(\zeta _2^2 + \zeta _3^2\big){\zeta _1}, \\ 
{\tau _0}\frac{{\partial {\zeta _2}}}{{\partial t}} = \mu {\zeta _2} + h{\zeta _1}{\zeta _3} - \sigma_1^\prime\zeta _2^3 - \sigma_2^\prime\big(\zeta _1^2 + \zeta _3^2\big){\zeta _2}, \\ 
{\tau _0}\frac{{\partial {\zeta _3}}}{{\partial t}} = \mu {\zeta _3} + h{\zeta _1}{\zeta _2} - \sigma_1^\prime\zeta _3^3 - \sigma_2^\prime\big(\zeta _1^2 + \zeta _2^2\big){\zeta _3}. 
\end{array}
\label{finalampli}
\end{equation}
System \eqref{finalampli} admits different kinds of solutions such as stationary states, stripe patterns, hexagonal patterns and mixed states. To investigate pattern formation, we perform a linear stability analysis. Depending on the parameters $\mu$, $\sigma _{1}^\prime$, $\sigma _{2}^\prime$ and $h$, different structural patterns can emerge in system \eqref{2Dmodel}. As the Jacobian matrix of system \eqref{finalampli} is given by
\begin{equation*}
	\left( {\begin{array}{*{20}{c}}
			{\mu - 3\sigma_1^\prime\zeta_1^2 - \sigma_2^\prime\big( {\zeta_2^2 + \zeta_3^2} \big)}&{\left| h \right|{\zeta _3} - 2\sigma_2^\prime{\zeta _1}{\zeta _2}}&{\left| h \right|{\zeta _2} - 2\sigma_2^\prime{\zeta _1}{\zeta _3}}\\
			{\left| h \right|{\zeta _3} - 2\sigma_2^\prime{\zeta _1}{\zeta _2}}&{\mu - 3\sigma_1^\prime\zeta_2^2 - \sigma_2^\prime\big( {\zeta_1^2 + \zeta_3^2} \big)}&{\left| h \right|{\zeta _1} - 2\sigma_2^\prime{\zeta _2}{\zeta _3}}\\
			{\left| h \right|{\zeta _2} - 2\sigma_2^\prime{\zeta _1}{\zeta _3}}&{\left| h \right|{\zeta _1} - 2\sigma_2^\prime{\zeta _2}{\zeta _3}}&{\mu - 3\sigma_1^\prime\zeta_3^2 - \sigma_2^\prime\big( {\zeta_1^2 + \zeta_2^2} \big)}
	\end{array}} \right),
\end{equation*}
we can identify the following cases:
\begin{itemize}
\item{The stationary state is given by ${\zeta _1} = {\zeta _2} = {\zeta _3} = 0$ and is stable for $\mu< 0$ and unstable for $\mu> 0$.}
\item{The hexagonal pattern exists when $\zeta = {\zeta _1} = {\zeta _2} = {\zeta _3} = \frac{{\left| h \right| \pm \sqrt {{h^2} + 4\left( {{\sigma _{1}^\prime} + 2{\sigma _{2}^\prime}} \right)\mu } }}{{2\left( {{\sigma _{1}^\prime} + 2{\sigma _{2}^\prime}} \right)}}$, with $\varphi = 0$ or $\varphi = \pi$ and $\mu > {\mu _1} = \frac{{ - {h^2}}}{{4\left( {{\sigma _{1}^\prime} + 2{\sigma _{2}^\prime}} \right)}}$. The hexagonal pattern $H_{\pi}$ (when $\varphi = \pi$) is stable for $\mu < {\mu _4} = \frac{{2{\sigma _{1}^\prime} + {\sigma _{2}^\prime}}}{{{{\left( {{\sigma _{2}^\prime} - {\sigma _{1}^\prime}} \right)}^2}}}{h^2}$ only, whereas, $H_0$ is always unstable, when $\varphi = 0$.}
\item{Finally, the mixed state exists when $\sigma _{2}^\prime > \sigma _{1}^\prime$ and $\mu > \sigma _{1}^\prime \zeta _1^2=\mu_3$, where ${\zeta _1} = \frac{|h|}{\sigma _{2}^\prime - \sigma _{1}^\prime}$, ${\zeta _2} = \sqrt{\frac{\mu - \sigma _{1}^\prime \zeta _1^2}{\sigma _{1}^\prime + \sigma _{2}^\prime}}$, ${\zeta _3} = \sqrt{\frac{\mu - \sigma _{1}^\prime \zeta _1^2}{\sigma _{1}^\prime + \sigma _{2}^\prime}}$, and is always unstable.}
\end{itemize}

\begin{figure}
\centering
\includegraphics[width=18cm,height=9.5cm]{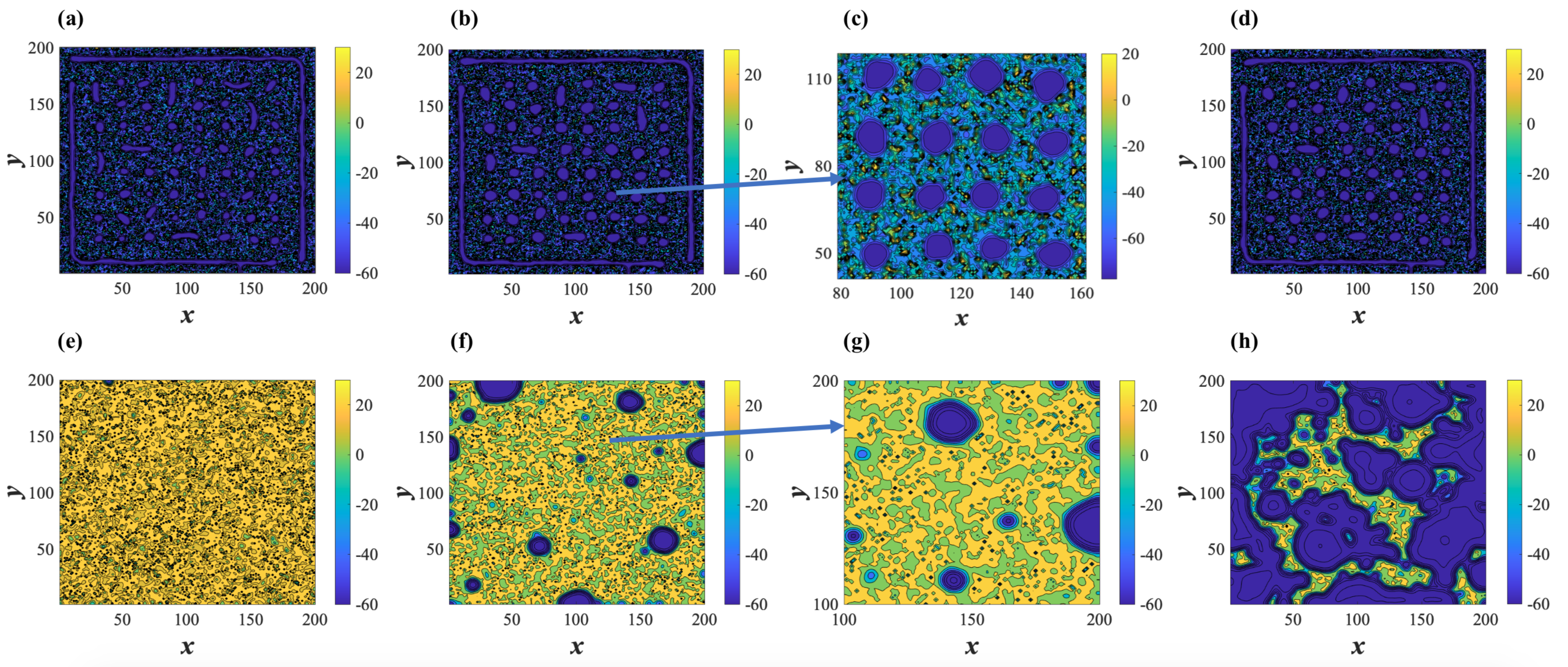}
\caption{Pattern formation in system \eqref{2Dmodel} of diffusively coupled, 2D Izhikevich neurons. First row of panels: set I for $I=-105.1$, $D_{11}=0.1$ and $D_{22}=9$: at (a)  $t=100$, (b)  $t=150$ and (d)  $t=200$. Second row of panels: set III for $I=80$, $D_{11}=1.46$ and $D_{22}=0.01$: at (e)  $t=60$, (f)  $t=63$ and (h)  $t=65$. Panels (c) and (g) show zoom-in versions of panels (b) and (f), respectively. Note that in all panels, the colorbar encodes the values of the solution $u(x,y)$ of system \eqref{2Dmodel}.}\label{pattern}
\end{figure}

In the next section, we present numerical results that corroborate the analytical results derived in this section using the amplitude equations.

\subsection{Numerical results for diffusively coupled Izhikevich neurons on a 2D spatial domain}

Here, we analyze the spatiotemporal evolution of patterns exhibited by system \eqref{2Dmodel} for different diffusion couplings. We consider an extended medium with 2D spatial diffusion and $N \times N$ coupled, equally spaced, identical Izhikevich neurons given by system \eqref{2Dmodel}, where $N=200$. We solve numerically system \eqref{2Dmodel} using the forward Euler method with a central difference representation, which is a finite-difference scheme, with step sizes $\Delta x =\Delta y = \Delta = 0.5$ and $\delta t = 0.001$ for the spatial mesh and time-integration step size, respectively. The initial conditions are considered with appropriate periodic perturbations as in the case of system \eqref{model}. We denote the variations in the membrane voltage of a single neuron at the spatial coordinate $(i, j)$ at time $t$ as $v_{i,j} (t)$. The partial derivative is then approximated by $\frac{{{\partial ^2}v}}{{\partial {x^2}}} + \frac{{{\partial ^2}v}}{{\partial {y^2}}} = \frac{1}{{{\Delta^2}}}\big( {{v_{i - 1,\,j}} + {v_{i + 1,\,j}} + {v_{i,\,j - 1}} + {v_{i,\,j + 1}} - 4{v_{i,\,j}}} \big)$. We note that here we want to investigate the characteristics of pattern formation in the context of the diffusive system \eqref{2Dmodel}, where the diffusion couplings account for the synaptic couplings among the Izhikevich neurons on the 2D spatial domain.

In the following, we consider positive diffusion couplings $D_{11}$ and $D_{22}$ such that a Turing instability can manifest in the dynamics of system \eqref{2Dmodel}. Then, we identify the conditions for the emergence and stability of stationary, hexagonal and mixed-state patterns, using the amplitude equations. The conditions for the emergence and stability properties of these spatiotemporal patterns depend on the parameter values of the system, among which are the diffusion couplings $D_{11}$ and $D_{22}$. Hence any pair of $D_{11}$ and $D_{22}$ values that satisfy the above conditions, could be a choice.

The spatial patterns observed are organized into simple hexagons and mixed-state arrangements of spot-like formations, which are temporarily stationary. The patterns that arise around the spatially uniform steady-state solution occur at a finite wave number. For $I=-105.1$, $D_{11}=0.1$ and $D_{22}=9$ of set I, we obtain $h=-0.3079$, $\sigma _{1}^\prime=1.0675$, $\sigma _{2}^\prime = 2.9011$, $\mu_1= -0.0034$, $\mu_3= 0.0301$, $\mu_4 = 0.1420$ and $\mu = 0.0106$. Clearly, $\mu >0$, which indicates the stationary state is unstable. Here, $\mu > \mu_1$ and $h<0$, which pinpoint to the existence of hexagonal patterns, $H_\pi$ \cite{iqbal2017pattern,zhao2014turing,yuan2013spatial,zhang2014spatio}. Also, since $\mu < \mu_4$, it confirms the existence of stable hexagons. We perform the linear stability analysis for hexagonal patterns and pay particular attention to the destabilizing modes that are moved with respect to the patterns.  Summarize the above analysis, we can show our results as a bifurcation diagram in Fig. \ref{dispersion}(e). The $H_\pi$ patterns are stable in the region $\mu < \mu_4$. The $H_0$ patterns are always unstable. We have validated the existence of spatial modulation of hexagonal patterns by solving Eq. \eqref{finalampli} and present them in Fig. \ref{pattern} (a)-(d) for long integration times. A zoom-in, shown in Fig. \ref{pattern}(c), confirms their appearance, which is shown in small patches. The hexagonal patterns, $H_\pi$, are observed on the whole domain. Interestingly, we have shown both the low-amplitude oscillations (deep-blue regions) and high amplitude oscillations (small yellow patches) to validate the time-independent hexagonal patterns. The numerical values for the existence of stable hexagons are given by $\zeta_1 = \zeta_2 = \zeta_3 = 0.000000071 \approx 0$, which are also verified by the analytical values discussed in Subsec. \ref{subesec_as}. The spatial instabilities lead to spatially modulated hexagons and other mixed states characterized by mixed-type patterns. Similarly, for $I=80$, $D_{11}=1.46$ and $D_{22}=0.01$ of set III, we find $h=0.0058$, $\sigma _{1}^\prime=-0.0002534$, $\sigma _{2}^\prime = -0.00023657$, $\mu_1= 0.0116$, $\mu_3= -30.2297$, $\mu_4 = -88.6813$ and $\mu = 0.0118$. Here,  the stationary state is unstable as $\mu>0$. The above numerical values confirm the existence of unstable mixed-type patterns (as $\sigma _{2}^\prime > \sigma _{1}^\prime$ and $\mu > \mu_3$) and unstable hexagons, $H_0$, (as $\mu > \mu_1$ and $h>0$), shown in Fig. \ref{pattern}(e)-(h).  However, as time increases, the patterns become distorted and mixed-state patterns emerge. The coexistence of ordered and disordered states are shown in Fig. \ref{pattern}(f). With further increase in time, most of the neurons' activity leads to a steady state, shown in Fig. \ref{pattern}(h). Finally, plots of the corresponding time-series are shown in Fig. \ref{dispersion}(c)-(d), using Eq.\eqref{finalampli}.

\section{Conclusions}

In this work, we introduced a mathematical methodology to analyze and characterize the propagation of wavy patterns in diffusively coupled Izhikevich neurons on a 2D spatial domain. Our approach focused on the development of a theory including diffusive, Turing instabilities. The non-uniform, spatiotemporal patterns exhibit collective dynamics that can be found in neural recordings for particular working mechanisms in brain functioning \cite{keane2015propagating,townsend2018detection,meier2015bursting,schiff2007dynamical,townsend2015emergence,mhatre2012grid,liu2015dynamical}.

First, we described the oscillatory behavior of a single Izhikevich neuron and presented the corresponding bifurcation analysis. Next, we explored diffusively coupled Izhikevich neurons on a 2D configuration. We characterized the spatial instability and derived a three-mode normal form. Later, this normal form was used to study pattern formation in the diffusive system. We explained theoretically and showed numerically the emergence of hexagonal patterns. Even for suitable parameter regimes at which stable hexagons exist, persistent periodic and irregular dynamical behavior can be found \cite{sain2000instabilities}.
We focused on instabilities that lead to the emergence of various patterns in the Turing domain. Sometimes, the oscillatory short-wave instabilities can form stable, periodically modulated, hexagons.
Furthermore, the spatial instabilities lead to spatially modulated hexagons and other mixed states characterized by mixed-type patterns. Consequently, we have been able to verify the observed spatiotemporal activities with our analytical method. 
Thus, our present study  and the work in \cite{mondal2019diffusion}, show clearly that the mathematical approach used here can be generalized to study analytically spatiotemporal instabilities and pattern formation in excitable extended neural models, including in Turing and non-Turing regimes.
The diffusively coupled system showed dynamical coherent structures and the single-neuron model showed highly variable complex firings. This may be useful in future work to investigate the spatial scale of neural features with various oscillatory signals as the low-amplitude oscillations affect the high-amplitude oscillations \cite{townsend2018detection}. Our theoretical approach explored some of the mechanisms of pattern formation pertinent to systems of spiking neurons \cite{keane2015propagating,song2018classification}. The observed hexagonal patterns might reflect the rhythmic activities of neural dynamics such as information processing in locally connected diffusive networks. The reported results have potential to contribute significantly to the study of diffusively-coupled neural systems in general.\\\\
{\bf Appendix}\label{appendixa}\\
\indent The coefficients of Eq. \eqref{4.12} are given by
$\left( {\begin{array}{*{20}{c}}
	{{V_0}}\\
	{{U_0}}
	\end{array}} \right) = \left( {\begin{array}{*{20}{c}}
	{{z_{v0}}}\\
	{{z_{u0}}}
	\end{array}} \right)\Big({\left| {{\Theta_1}} \right|^2} + {\left| {{\Theta_2}} \right|^2} + {\left| {{\Theta_3}} \right|^2}\Big),\,\,{V_j} = {f}{U_j}$,
$\left( {\begin{array}{*{20}{c}}
	{{V_{jj}}}\\
	{{U_{jj}}}
	\end{array}} \right) = \left( {\begin{array}{*{20}{c}}
	{{z_{v1}}}\\
	{{z_{u1}}}
	\end{array}} \right)\Theta_j^2,\,\,\left( {\begin{array}{*{20}{c}}
	{{V_{jk}}}\\
	{{U_{jk}}}
	\end{array}} \right) = \left( {\begin{array}{*{20}{c}}
	{{z_{v2}}}\\
	{{z_{u2}}}
	\end{array}} \right){\Theta_j}{\bar \Theta_k}$,
where
${z_{v0}} = \frac{2{a_{22}^T}h_1}{A},\,\,{z_{u0}} = \frac{-2{a_{21}^T}h_1}{A}$,
${z_{v1}} = \frac{\big({a_{22}^T}-4D_{22}k_T^2\big)h_1}{B},\,\,{z_{u1}} = \frac{-{a_{21}^T}h_1}{B}$,
${z_{v2}} = \frac{2\big({a_{22}^T}-3D_{22}k_T^2\big)h_1}{C},\,\,{z_{u2}} = \frac{-2{a_{21}^T}h_1}{C}$,
$A = {a_{11}^T}{a_{22}^T} - {a_{12}^T}{a_{21}^T} \mbox{ and }
B = \Big({a_{11}^T}-4D_{11}k_T^2\Big)\Big({a_{22}^T}-4D_{22}k_T^2\Big) - {a_{12}^T}{a_{21}^T},
C = \Big({a_{11}^T}-3D_{11}k_T^2\Big)\Big({a_{22}^T}-3D_{22}k_T^2\Big) - {a_{12}^T}{a_{21}^T}$.
The coefficients of Eq. \eqref{4.13} are given by
$H = 0.08{f}^2,\,\,{\sigma _1} = -0.08(z_{v0}+z_{v1})f$ and ${\sigma _2} = -0.08(z_{v0}+z_{v2})f$.
Finally, the coefficients of Eq. \eqref{ampli} are given by
${\tau _0} = \frac{f+g}{{{I_T}{m_{11}}{f}}},\,\,\mu = \frac{{I - I_T}}{{{I_T}}},\,\,h = \frac{{H}}{{{I_T}{m_{11}}{f}}},\,\,\sigma_1^\prime = \frac{{{\sigma _1}}}{{{I_T}{m_{11}}{f}}}$ and $\sigma_2^\prime = \frac{{{\sigma _2}}}{{{I_T}{m_{11}}{f}}}$.
 
\renewcommand{\baselinestretch}{0.01}
\renewcommand*{\bibfont}{\small}

\end{document}